\documentclass[prl,aps,reprint,citeautoscript,amsmath,amssymb,superscriptaddress]{revtex4-1}
\usepackage{graphicx}% Include figure files
\usepackage{dcolumn}% Align table columns on decimal point
\usepackage{bm}% bold math
\usepackage{amsmath}
\usepackage{hyperref} %hyper references
\usepackage{soul} % use this to cross out a text
\DeclareGraphicsExtensions{.pdf,.eps,.png,.jpg,.mps}
\usepackage{mathrsfs}
\usepackage{soul} %Para tachar escritura
\usepackage[utf8]{inputenc} %tildes
\usepackage{hyperref} %hyper references
\usepackage{xcolor} %color

\usepackage{booktabs}

\begin{document}
\title{Reversible edge spin currents in antiferromagnetically proximitized dichalcogenides}

\author{Natalia Cortés}
\email{natalia.cortesm@usm.cl}
\affiliation{Departamento de Física, Universidad Técnica Federico Santa María, Casilla 110V, Valparaíso, Chile}
\affiliation{Department of Physics and Astronomy, and Nanoscale and Quantum Phenomena Institute, Ohio University, Athens, Ohio 45701–2979, USA}
\author{O. Ávalos-Ovando}
\email{Current address: Department of Physics, Ohio State University, Columbus, Ohio 43210, USA}
\affiliation{Department of Physics and Astronomy, and Nanoscale and Quantum Phenomena Institute, Ohio University, Athens, Ohio 45701–2979, USA}
\author{S. E. Ulloa}
\affiliation{Department of Physics and Astronomy, and Nanoscale and Quantum Phenomena Institute, Ohio University, Athens, Ohio 45701–2979, USA}

\date{\today}

%***********************************************************************
\begin{abstract}
We explore proximity effects on transition metal dichalcogenide ribbons deposited on antiferromagnetic (AFM) insulating substrates. We model these hybrid heterostructures using a tight-binding model that incorporates exchange and Rashba fields induced by proximity to the AFM material. The robust edge states that disperse in the midgap of the dichalcogenide are strongly affected by induced exchange fields that reflect different AFM ordering in the substrate.  This results in enhanced spin-orbit coupling effects and complex spin projection content for states on zigzag ribbon edges.  Gated systems that shift the Fermi level in the midgap range are also shown to exhibit spin polarized currents on these edges.  Antiparallel exchange fields along the edge results in spin currents that can reverse polarization with the applied field. The added functionality of these hybrid structures can provide spintronic devices and versatile platforms to further exploit proximity effects in diverse material systems.  

\end{abstract}

%\keywords{XX\sep XX \sep XX}
%\pacs{XX}

\maketitle

\emph{Introduction}.-- Proximity coupling of two-dimensional (2D) materials with a variety of substrates is an emergent research field greatly expanding the properties of the different constituents, and creating new optical, transport and even magnetic functionalities  \cite{vzutic2019proximitized,Gongeaav4450,NNRev2019}. Theory and experiments have demonstrated that proximity-induced wave function overlap can alter the electronic response, as evidenced in the spectra of diverse heterostructures \cite{yang2013proximity,qiao2014quantum,qi2015giant,Zhang2016,wei2016strong,leutenantsmeyer2016proximity,su2017effect}. 
Prominent among many interesting substrates are those with antiferromagnetic (AFM) order. They possess zero net magnetization, are resistant to stray fields and have faster dynamics (in the terahertz range) than most ferromagnetic (FM) crystals, which makes them suitable candidates for magnetoelectronic and novel spintronics applications \cite{jungwirth2016antiferromagnetic,jungwirth2018multiple,gomonay2018antiferromagnetic,li2020spin,Baltz2018}.

The strong spin-orbit coupling (SOC) in two-dimensional semiconducting transition metal dichalcogenides (TMDs) couples spin and valley degrees of freedom \cite{Xiao2012,liu2015electronic,xu2014spin} at the direct bandgap valleys $K$ and $K'$ in the Brillouin zone \cite{Mak2010,wang2012electronics}. In close proximity to ferromagnets \cite{Zhao2017,Zhong2017van,Seyler2018,zou2018probing,Ciorciaro2020}, the valley degeneracy of the TMD monolayer is lifted due to the broken time reversal symmetry (TRS), producing spin-splittings in the bands because of the competition between SOC and the induced exchange field. An enhanced valley splitting sensitivity to applied fields of up to 16 meV/T has been measured in a 2D WS$_2$-EuS system \cite{norden2019giant}, a giant valley splitting (300 meV) and sizable Rashba field ($\lesssim 100$ meV) are predicted for the commensurate MoTe$_2$-EuO heterostructure \cite{qi2015giant,Zhang2016}, and giant valley splittings in WS$_2$-VN \cite{Ke2019}.   
 
The proximity effect on TMD deposited on insulating AFM substrates has been recently explored theoretically in different structures \cite{Xue2019}, including WS$_2$-MnO(111) \cite{xu2018large}, MoS$_2$-CoO(111) \cite{yang2018induced}, TMD-CrI$_3$(bilayer) \cite{zollner2019proximity} and MoTe$_2$-RbMnCl$_3$(001) \cite{li2018large}. In these hybrid systems, the atoms of the topmost layer of the substrate closer to the 2D material are typically ferromagnetically coupled, so that the exchange fields on the TMD break TRS and generate valley splittings ranging from few to hundreds of meV, similar to the case of TMD on FM substrates. The obvious question that arises is if a full AFM ordering induced by proximity may contribute to richer spin-dependent electronic structure and more complex spin dynamics in TMD-AFM hybrids \cite{gomonay2014spintronics,jungwirth2018multiple,gomonay2018antiferromagnetic}. We explore this issue here theoretically, focusing on TMD nanoribbons with zigzag edges in order to explore how the spin-valley structure and robust structural protection of the material result in interesting behavior and functionalities. Spin-orbit torques are predicted to depend on the sublattice symmetry potential in AFM honeycomb lattices with Rashba SOC \cite{sokolewicz2019spin}, for example. 

Zigzag-terminated ribbons possess asymmetric edges with intrinsic Rashba SOC \cite{rostami2016edge,avalosEdges2016,shahbazi2018linear}. Robust edge-states have been shown to be stable on prepatterned substrates \cite{Cheng2017nanoletters}, and spin-polarized currents are predicted when the zigzag ribbons are in proximity to FM substrates \cite{cortes2019tunable}. Other stable one-dimensional (1D) conducting channels have been demonstrated at twin-grain boundaries in MoTe$_2$ \cite{diaz2016high}, and Majorana bound states at the ends of zigzag edges have been proposed for dichalcogenide nanoribbons in proximity to a superconductor \cite{Chu2014}. Control of edge currents is one of the most sought-after quantities in van der Waals engineering, and the 1D spin-polarized edge channels involving proximity effects here provide an ideal tunable platform. 

We present here a study of MoTe$_2$-AFM hybrid heterostructures with different AFM patterns that can induce both magnetic exchange and Rashba fields on the TMD nanoribbons. The hybrids are modeled by means of a three-orbital tight-binding  Hamiltonian \cite{Liu2013}, which includes proximity-induced terms, allowing us to describe the spin-dependent electronic states and associated spin dynamics, especially for the edge modes in the 2D bulk gap. The proximity effects result in augmented Rashba SOC and spin-splitting for the edge modes in the TMD ribbon.  Although the overall order is AFM, we find that if the induced moments are parallel along the zigzag edge of the TMD, the system can carry spin-polarized currents whenever the Fermi level is shifted to the bulk midgap, similar to the case when on a ferromagnetic substrate.  However, when the induced moments are antiparallel along the ribbon edges, the spin currents can be tuned in magnitude and be fully reverse direction.  This interesting reversibility promises additional functionalities and spintronic devices.

%\section{System}\label{sec:SystemAFM}

\begin{figure}[!h]
\centering
\includegraphics[width=1.0\linewidth]{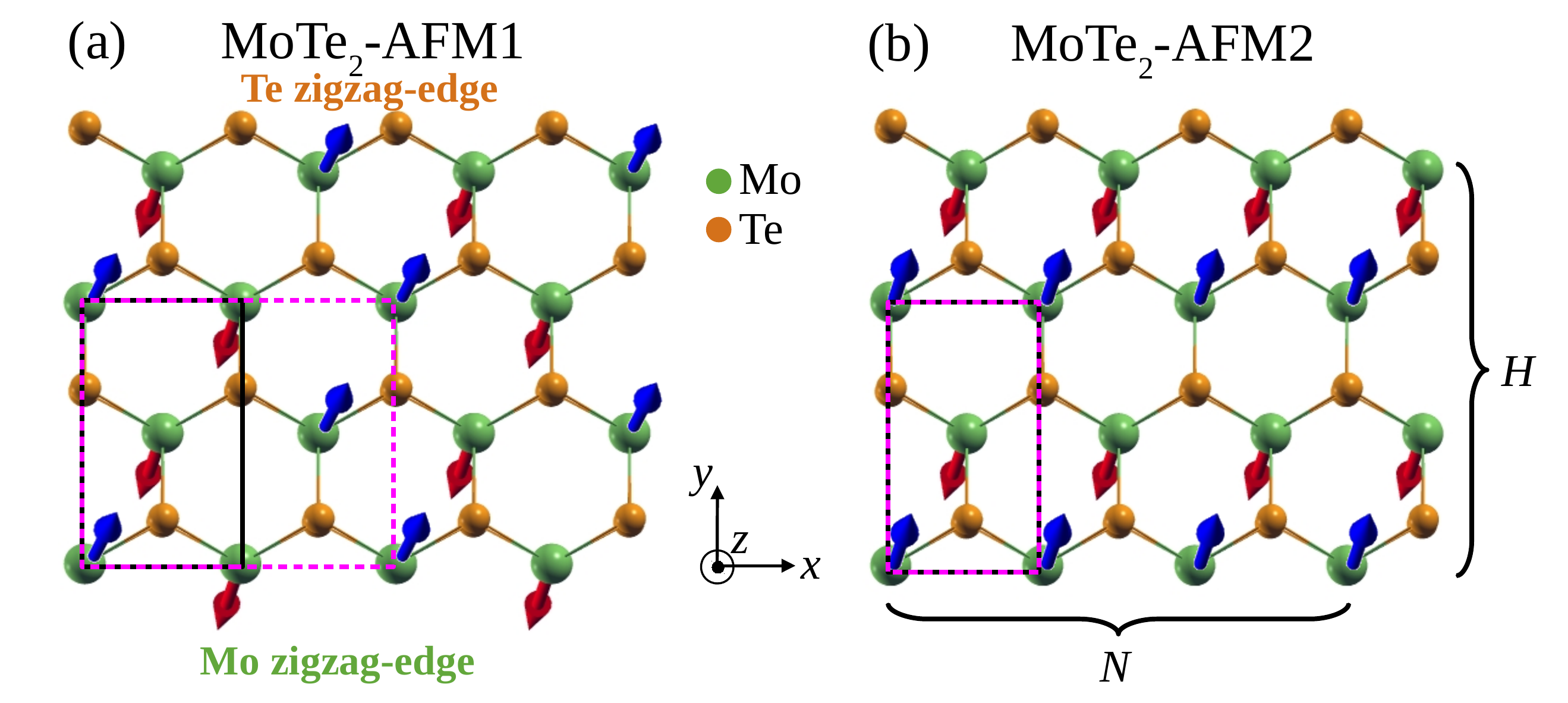}
\caption{MoTe$_2$ ribbons in proximity to AFM substrates with different spatial arrangement of induced exchange fields, (a) AFM1, and (b) AFM2. AFM substrates are not shown; ribbon size and shape are schematic, with length $N$, width $H$, and zigzag edges along the $x$-axis. The Mo (Te)-zigzag edge is at $y=0$ $(H)$. 
Top views (slightly tilted) of projected moments on the Mo atoms shown as blue (up) and red (down) arrows. Magenta (black) rectangles indicate magnetic (geometric) unit cell, with periodic boundary conditions along the $x$-direction.}\label{fig1}
\end{figure}

\emph{System}.--- 
To construct the TMD-AFM heterostructures, the selection of possible AFM substrates is large and includes Mn-based compounds MnPX$_3$ ($\mathrm{X=S, Se}$), which have hexagonal lattice, and are semiconducting \cite{li2013coupling}, as well as MnO \cite{xu2018large} or CoO \cite{yang2018induced}.  For concreteness we study here MoTe$_2$ ribbons, and depending on the specific material and crystalline face proximate to the TMD, one could have different underlying AFM ordering. We have considered two types of AFM order, labeled AFM1 and AFM2, both assumed commensurate with the TMD lattice and shown in Fig.\ \ref{fig1}. The induced exchange fields on the TMD are shown as antiparallel moments with opposite magnetization and different periodic patterns from the AFM substrate projected onto the transition metal sublattice, and represented by blue (up) and red (down) arrows. In the MoTe$_{2}$-AFM1 heterostructure shown in Fig.\ \ref{fig1}(a), the magnetic unit cell is twice as large as the lattice unit cell and the induced magnetization alternates along the horizontal zigzag edges. For the MoTe$_{2}$-AFM2 in Fig.\ \ref{fig1}(b), the magnetic and lattice cells are the same, as adjacent rows with the same magnetization alternate across the ribbon but result in parallel moments along the zigzag edges.
Ribbons have length $N$ and width $H$, with $N$ even for the AFM1 substrate, to enforce net zero magnetization. When on the AFM2 substrate, $H$ must be similarly even. We typically consider ribbons with 1600 Mo-sites with width of $\sim$125 \AA{} (40 Mo sites), and length of $\sim$144 \AA{} (40 Mo sites), with periodic boundary conditions along the zigzag edges. 

%

%%%%%%%%%%%%%%%%%%%%%%%%%%%%%%%%%%%%%%%%%%%%%%%%%%%%%%%%%%%%
\emph{Tight-binding description}.--The effects of the AFM substrate are incorporated into the MoTe$_{2}$ lattice Hamiltonian by onsite magnetic exchange and Rashba fields,
\begin{equation}\label{full3OTBAFM}
\mathcal{H}_{\mathrm{MoTe}_{2}\rm{-}\mathrm{AFM}}= \mathcal{H}_{\mathrm{MoTe}_{2}}+\mathcal{H}_{\mathrm{ex},j}+\mathcal{H}_\mathrm{R}.
\end{equation}
$\mathcal{H}_{\mathrm{MoTe}_{2}}$ describes the pristine TMD Hamiltonian, including intrinsic SOC and a matrix of hoppings up to first-neighbors. It is written in a basis of relevant transition metal $d$-orbitals, $\big\{ \left|d_{z^{2}},s \right\rangle$, $\left|d_{xy},s \right\rangle$, $\left|d_{x^{2}-y^{2}},s \right\rangle \big\}$, with spin index  $s=\uparrow,\downarrow$ \cite{Liu2013}. Unlike an FM substrate in proximity to the TMD monolayer, where the induced magnetic exchange field has the same contribution on all sites \cite{qi2015giant,Zhang2016,cortes2019tunable}, here the exchange Hamiltonian $\mathcal{H}_{\mathrm{ex},j}$ is site-dependent, with opposite moments depending on the expressed AFM periodicity \cite{supplemental}. We use $B_{c}=206$ meV and $B_{v}=170$ meV as typical values for the conduction and valence band induced exchange fields, respectively \cite{qi2015giant}.
The broken symmetry generated by the proximity with the AFM substrate also generates an interfacial Rashba field \cite{kane2005quantum,ochoa2013spin,qi2015giant,frank2018,cortes2019tunable}, mixing the spin and orbital components in the MoTe$_{2}$ monolayer, with coupling $\lambda_{R}$, estimated to be few tens of meV. The Rashba field provides an overall canting of spins in the monolayer bulk, as it competes with the exchange field. As we will see, it modifies the intrinsic SOC along the edge modes of freestanding zigzag TMD ribbons \cite{Chu2014}. All parameters are estimated/taken from density functional calculations that consider magnetic substrates, where exchange fields are found to exhibit similar values when the TMD monolayer is in proximity to either an FM or AFM substrate \cite{qi2015giant,zollner2019proximity}, and the atoms closer to the TMD dominate.
%
%%%%%%%%%%%%%%%%
\begin{figure*}[!ht]
\centering
\includegraphics[width=1\linewidth]{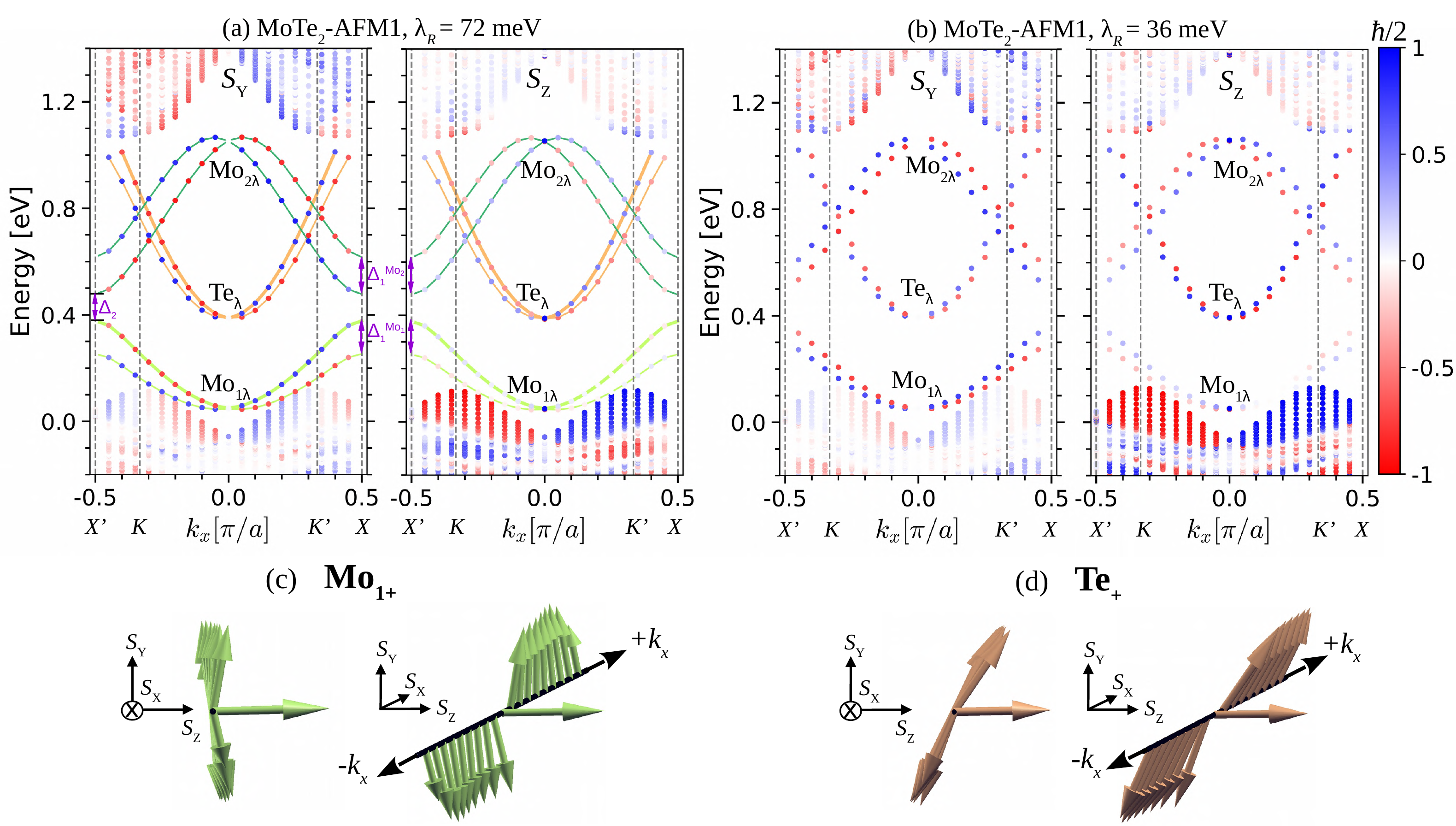}
\caption{Spin projections of the energy spectrum for zigzag ribbon in MoTe$_2$-AFM1 system. Mo$_{1\lambda}$, Mo$_{2\lambda}$ and Te$_\lambda$ label the edge modes in the bulk midgap, indicated with light and dark green, and orange solid lines, respectively. In (a) the interfacial Rashba field $\lambda_\text{R}=72$ meV; in (b) is set to half that value. Left panels in (a) and (b) show $S_{\text{Y}}$ projection; right panels show $S_{\text{Z}}$. Color bar indicates positive (negative) spin projection as blue (red) gradient. $\Delta_{1}^{\text{Mo}_l}$ and $\Delta_2$ indicate the spin-splitting and intramode gap, respectively, at the $X, X'$ valleys.  Panels (c) and (d) illustrate spin projections for different momentum $X'\leq k_{x} \leq X$ for both higher Mo$_{1,+}$ and Te$_+$ branches, highlighted with thicker light green and orange lines respectively in panel (a). The arrow’s size (direction) indicates the magnitude (orientation) of the spin projection \cite{externalfield}.}\label{fig2}
\end{figure*}
%%%%%%%%%%%%%%%%%%%%%%%%%%%%%%%

\emph{Proximitized AFM zigzag edges}.---Figure \ref{fig2}(a) shows the spin projections $S_{\text{Y}}$ and $S_{\text{Z}}$ for the electronic spectrum of a MoTe$_2$-AFM1 heterostructure with typical interfacial Rashba field $\lambda_{\text{R}}=72$ meV, for momentum $k_x$ along the long ribbon direction. 
The $S_{\text{X}}$ component is vanishingly small, as the Rashba field here is along the $y$-axis \cite{externalfield}. For comparison, Fig.\ \ref{fig2}(b) shows results for a weaker interfacial Rashba field $\lambda_{\text{R}}=36$ meV \cite{expRashbafield}. The combined effect of the AFM1 exchange and Rashba fields leads to antisymmetric dispersions about $k_x=0$ for the edge modes dispersing through the bulk gap. Contrary to ribbons (or monolayers) on FM substrates \cite{qi2015giant,cortes2019tunable}, the spectrum here shows zero valley polarization in the bulk, preserving TRS in the structure, as the net induced magnetization is zero.  In the bulk bandgap region we can identify three pairs of dispersing edge modes, with wavefunctions localized on either the Mo edge, labeled Mo$_{1\lambda}$ and Mo$_{2\lambda}$, or on the Te edge, Te$_\lambda$ \cite{supplemental}. Each edge mode has two branches with eigenenergies $E_{\pm}^{\beta}(k_x)$ ($\beta=\text{Mo}_{1}, \text{Mo}_{2}, \text{Te}$), shown with solid lines in Fig.\ \ref{fig2}(a), and exhibiting spin orientation that depends strongly on $k_x$ and the corresponding $\lambda_{\text{R}}$. At the $\Gamma$ point, the edge states remain degenerate, as expected.  
%%%%%%%%%%%%%%%%%%%%%%%%%%%%%%%%%%%%%%%%%%%%%%%%%%%

The spin-splitting $\Delta_1^{\beta}(k_x)=E_{+}^{\beta}-E_{-}^{\beta}$ varies almost linearly with momentum $|k_x|$, and reaches its maximum value near the $X, X'$ points \cite{supplemental}, proportional to the interfacial Rashba couplings. An inter-mode gap $\Delta_{2}=E_{-}^{\text{Mo}_{2}}-E_{+}^{\text{Mo}_{1}}$ is present at the zone edges [see Fig.\ \ref{fig2}(a)] due to the double magnetic period induced by the AFM1 substrate. Notice that $\Delta_{2}$ increases with smaller $\lambda_{\text{R}}$, as the exchange field remains--see Fig.\ \ref{fig2}(b).   
The competition between SOC and induced exchange fields produces interesting spin structure content in the midgap edge modes, as mentioned above.  Figure \ref{fig2}(c) and (d) illustrate this for the system in panel \ref{fig2}(a), where the $S_{\text{Y}}$ projection is seen to dominate for all $k_x$, although $S_{\text{Z}}$ is non-zero.  This results in canted spin orientation for different $k_x$-energy values, which would prove important in the resulting edge currents to be discussed below.  Notice that smaller $\lambda_{\text{R}}$ values result in larger $S_{\text{Z}}$ projections, as seen in Fig.\ \ref{fig2}(b), and similar spin textures for the edge states.

The nanoribbon also sustains a Te$_\lambda$ edge mode only, appearing as a pair of higher energy branches that hybridize with the conduction bulk bands for $|k_x|\simeq \pi/2a$ in Fig.\ \ref{fig2}(a) [while not in \ref{fig2}(b)]. The spin projections for the Te$_{\lambda}$ modes are similar to those of the Mo$_{1\lambda}$ modes at the same $k_x$, as illustrated in Fig.\ \ref{fig2}(d).  The supplement shows examples of edge wavefunctions for the different modes \cite{supplemental}. 

\begin{figure}[!h]
\centering
\includegraphics[width=1\linewidth]{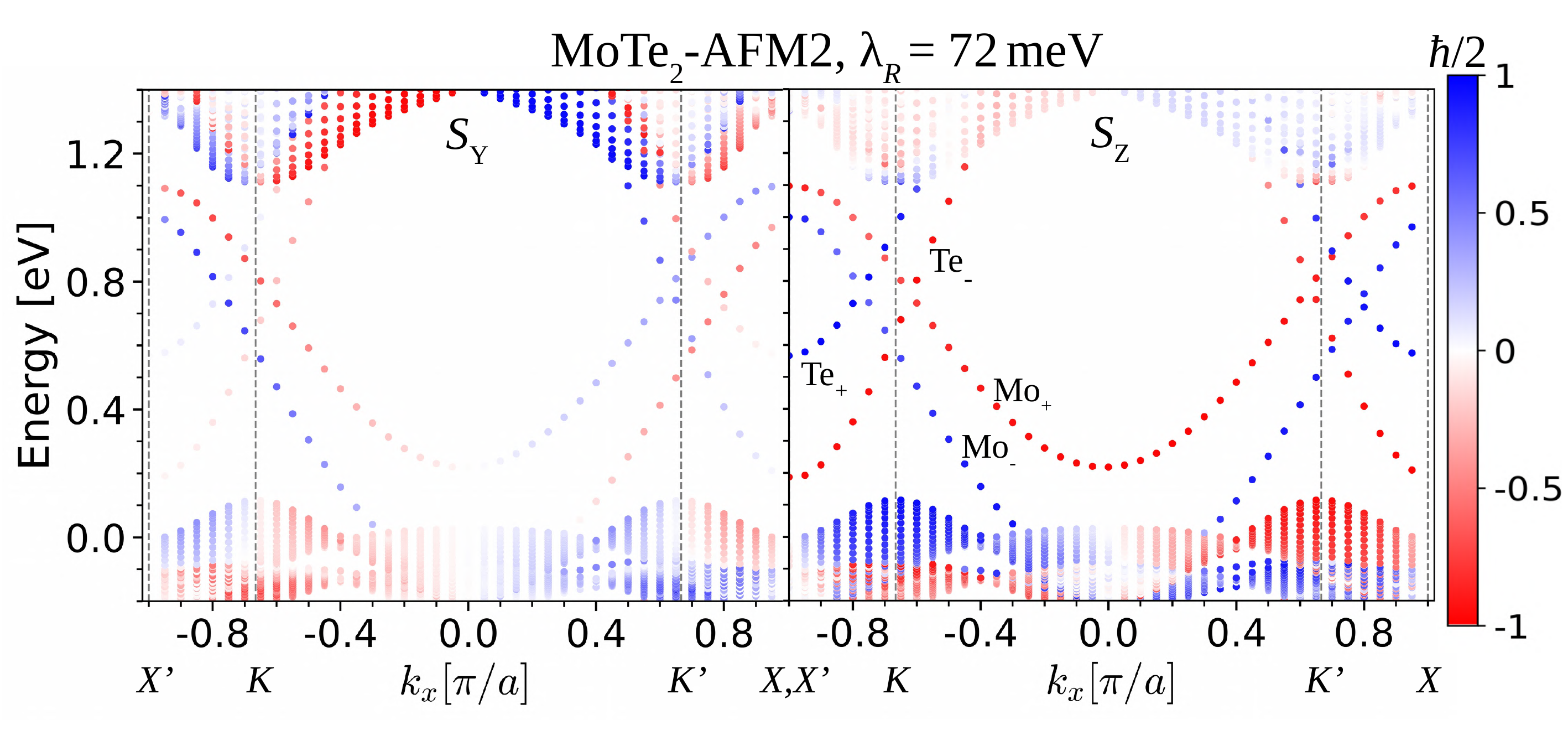}
\caption{$S_{\text{Y}}$ and $S_{\text{Z}}$ spin projections of the energy spectrum in the MoTe$_2$-AFM2 system in Fig.\ \ref{fig1}(b). Color bar indicates positive (negative) spin projection as blue (red) gradient. Nearly full $S_{\text{Z}}$ polarization for all midgap edge modes is due to parallel exchange fields along the zigzag edges of the ribbon.}\label{fig3}
\end{figure}

Another interesting system is the case of the MoTe$_2$-AFM2 heterostructure in  Fig.\ \ref{fig1}(b), which has parallel induced moments along the zigzag edges but opposite magnetization across the ribbon. The exchange field pattern in this case contributes as AFM coupling to the bulk, while being effectively FM along the zigzag edges. As a result, the induced moments preserve TRS in the bulk but break it for the electronic edge states. The energy spectrum is shown in Fig.\ \ref{fig3} with $S_{\text{Y}}$ and $S_{\text{Z}}$ projections, having clear differences with respect to the MoTe$_2$-AFM1 system. The broken TRS on the edges here produces strongly spin-polarized modes dispersing through the bulk midgap with large $S_{\text{Z}}$ projection, non-degenerate for $k_x=0$, and very similar behavior to the case when the nanoribbon is proximitized by a FM substrate \cite{cortes2019tunable}. Notice that upward dispersing modes are localized on the metallic Mo edge, as labeled.
If the exchange field is reversed, the edge modes reverse $S_{\text{Z}}$ as expected, with $S_{\text{Y}}$ determined by the interfacial Rashba field. Because the induced exchange on the zigzag edges behaves as an effective FM coupling, we can expect spin-polarized currents similar to the  MoTe$_2$-FM case \cite{cortes2019tunable}, propagating along the edges when the Fermi level lies in the bulk gap of the MoTe$_2$-AFM2 hybrid.  
%%%%%%%%%%%%%%%%%%%
%%%%%%%%%%%%%%%%%%
%\section{Spin Currents}

\emph{Spin currents}.--We now turn our attention to spin currents in these ribbons. As the system is gated and the Fermi energy is midgap, the current is only along the edges. 
We focus on the spin currents along the Mo edge in the MoTe$_2$-AFM1 structure of Fig.\ \ref{fig2}(a), as they allow more tunable spin polarization in the gap region.
For convenience, we fit the tight-binding dispersion and spin projections to an analytical next-nearest-neighbors tight-binding model with an appropriate AFM basis--see \cite{supplemental} for details. 

\begin{figure}[!h]
\centering
\includegraphics[width=1\linewidth]{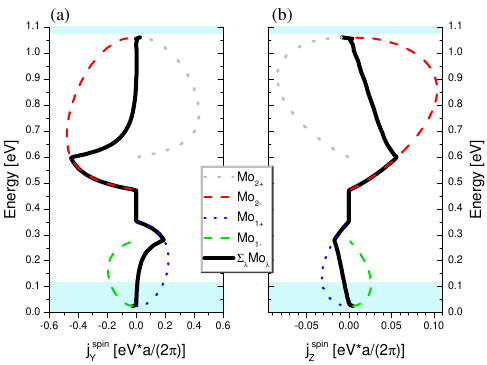}
\caption{Spin currents (horizontal axis) for components (a) $j_{\text{Y}}^{{spin}}$ and (b) $j_{\text{Z}}^{{spin}}$ along the Mo edge of the MoTe$_2$-AFM1 system of Fig.\ \ref{fig1}(a), for Fermi energies across the midgap (vertical axis)--blue shaded areas indicate bulk bands. Contributions of each Mo$_{l\lambda}$ branch are shown as dashed (dotted) lines, while total spin current is shown by solid black line. Notice different horizontal scales.}\label{fig4}
\end{figure}

The spin currents are given by $j_i^{spin}=S_{i}\frac{\partial E(k)}{\hbar\partial k}$ \cite{sinova2004universal}, where $i=\text{X,Y,Z}$. Figure \ref{fig4} shows $j_{\text{Y}}^{spin}$ [panel (a)] and $j_\text{Z}^{spin}$ [panel (b)] ($j_{\text{X}}^{spin}$ is negligible), for right movers with energies within the gap. The contribution from each Mo-branch is shown as dashed (Mo$_{l-}$) and dotted (Mo$_{l+}$) lines, with $l=1,2$, while the total spin current along the Mo-edge is shown by a solid black line. The spin current is non-zero mostly in the range between 0.2 eV and 0.7 eV, following the splitting of the different Mo-branches, and typically $j_\text{Y}^{spin} \simeq -10 j_\text{Z}^{spin}$ (notice different horizontal scales). Below 0.35 eV, only the Mo$_{1\lambda}$ bands compete with each other, reaching a maximum spin current at $E_F \simeq0.28$ eV, where only Mo$_{1+}$ contributes. A similar maximum value of $j_\text{Y}^{spin}$ is seen at $E_F \simeq0.6$ eV where the spin current is mostly contributed by Mo$_{2-}$. There is clearly no spin current in the $\Delta_2$ region where no propagating states exist. For energies $>0.6$ eV the competition between the Mo$_{2\lambda}$ bands decreases the spin current magnitude, until the Fermi energy reaches the conduction bulk bands and the current vanishes.

These results show that the spin current is highly tunable along the Mo-edge. As the Fermi level is shifted by gate fields, one would have the ability of selecting either a mostly \emph{up} $j_\text{Y}^{spin}$ current,  $E_{\text{F}}\lesssim 0.35$ eV, \emph{down} $j_\text{Y}^{spin}$ current,  $0.5\text{ eV}\lesssim E_{\text{F}} \lesssim 1.0\text{ eV}$, and corresponding smaller $j_\text{Z}^{spin}$ component, or no current at all for $E_{\text{F}}\in\Delta_2$, all within the same MoTe$_2$-AFM1 nanoribbon sample. This is in striking contrast to the MoTe$_2$-AFM2 and MoTe$_2$-FM cases \cite{cortes2019tunable}, where the Mo-edge will present spin currents with  varying magnitude but the same sign throughout the gap. We should further remark that one can anticipate that point defects will not affect the results significantly \cite{Chu2014,ridolfi2017electronic}. Realistic random bulk and edge disorder has been proved to only slightly suppress the edge currents in the MoTe$_2$-FM case, as the structural protection of the materials results in robust edge states  \cite{cortes2019tunable}. The reversibility of the spin current on AFM substrates and the robustness of edge states in these systems promises interesting "ambipolar" spin transistor devices within the same structure.

%\section{Conclusion}
\emph{Conclusions}.--Zigzag ribbons deposited on AFM substrates show midgap states with complex dispersion and associated spin current response that depend on the induced exchange field patterns. The proximity interaction with antiparallel induced moments along the zigzag edge of ribbons results in controllable \textit{up} and \textit{down} spin currents for midgap Fermi levels over wide Fermi energy range, where the Rashba spin-split Mo branches compete with each other. 
In contrast, the results for parallel induced moments are similar to those on proximitized ferromagnetic substrates, where spin-polarized currents along the edges for Fermi energy in the bulk gap exist, with tunable orientation and magnitude. The proposed hybrid systems here can serve as versatile platform to explore induced antiferromagnetism and 1D systems with complex spin textures.  These structures could be used to build novel devices in diverse solid-state operations such as tunable spin transistors.

%\section{Acknowledgments}
{\em Acknowledgments}.--N.C. acknowledges support from Conicyt, Fondecyt grant no.\ 3200658. O.\'A.-O. and S.E.U. acknowledge support from NQPI and NSF grant no.\ DMR 1508325. 

\bibliography{bib}

\begin{thebibliography}{51}
\bibitem [{ext()}]{externalfield}%
  \BibitemOpen
  \href@noop {} {}\bibinfo {howpublished} {Note that a small external field of
  1 meV along $S_\text{Z}$ is used to break the spin degeneracy, not affecting
  the overall results.}\BibitemShut {Stop}%
\end{thebibliography}
%\email{natalia.cortesm@usm.cl}

\end{document}